# Exploring the Statistical Properties of Outputs from a Process Inspired by Geometrical Interpretation of Newton's Method

Taki-eddine KIROUANI [*]

[*] Faculty of Exact Sciences and Natural and Life sciences, Echahid Cheikh Larbi Tebessi University, Algeria

Mail: ktaki@hotmail.fr

*Abstract:*

In this paper, the statistical properties of Newton's method algorithm output in a specific case have been studied. The relative frequency density of this sample converges to a well-defined function, prompting us to explore its distribution. Through rigorous mathematical proof, we demonstrate that the probability density function follows a Cauchy distribution. Additionally, a new method to generate a uniform distribution is proposed. To further confirm our findings, we employed statistical tests, including the Kolmogorov-Smirnov test and Anderson-Darling test, which showed high p-values. Furthermore, we show that the distribution of the distance between two successive outputs can be obtained through a transformation method applied to the Cauchy distribution.

## 1. Introduction

Newton's method is a powerful iterative technique for finding approximate solutions to real-valued functions [01]. It is especially useful for solving equations where analytical solutions are difficult or impossible to obtain. The Newton method has been known in mathematics for more than three hundred years, with its origins tracing back to the works of Sir Isaac Newton and Joseph Raphson [02].

One of the key advantages of Newton's method is its rapid convergence properties, particularly when the initial guess is close to the true root. The method's efficiency makes it a staple in numerical analysis and various applied fields, including physics, engineering, and economics. However, it also has limitations, such as the necessity for the derivative of the function and the potential for divergence if the initial guess is far from the root or if the function behaves poorly.

The idea behind this method is to create a sequence $x_{k+1} = g(x_k)$, where $g(x) = x - f(x)/f'(x)$ (01), according to the fixed-point theorem [03] and under certain verified conditions, the preceding sequence is expected to converge to the zeros of *f(x)*.

In this paper, we do not focus on the convergence or divergence behavior of this method. Instead, we study the statistical properties outputs generated by Newton's method Algorithm in a specific case.

## 2. An iterative process inspired by geometrical interpretation of Newton's method

Let's $P_2$ be the vector space of polynomials of degree of two or less, let's a set $F \subseteq P_2$, this set contains all polynomials of degree two that have no roots

$$F = \{p_2(x) = ax^2 + bx + c;\ a, c \in R^*, b \in R\ |b^2 - 4ac < 0\}$$

We can re-write $p_2(x)$ in the following form $p_2(x) = a((x - \bar{x})^2 + s^2)$ (02)

Such that $\bar{x} = -b/2a,\ s = \sqrt{D/4a^2},\ D = 4ac - b^2$

Take any polynomial $p_2(x)$ from set $F$ and let's define the following process inspired by geometrical interpretation of Newton's method

1- Take any value $x_1 \in \mathbb{R}^*$.
2- Draw the tangent line of curve $p_2(x)$ at point$(x_1, p_2(x_1))$.
3- Find the abscissa intersection point between the tangent line and $x$-axis.

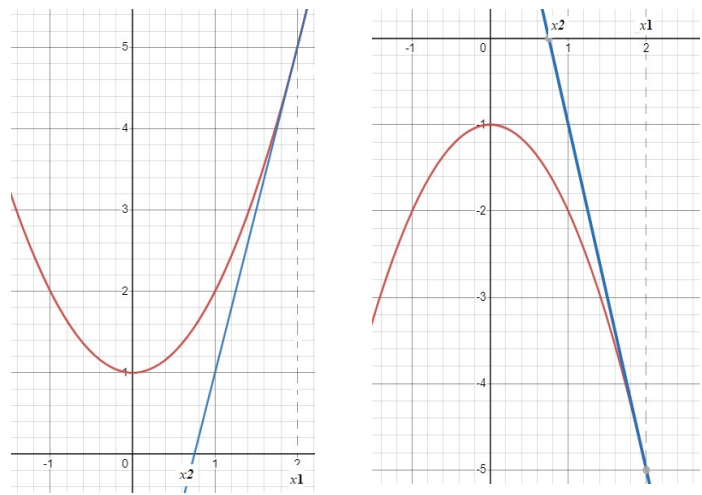

Case when $p(x) > 0$        Case when $p(x) < 0$

**Figure 1**. Plotting the parabola and tangent line for two cases

Repeat the previous steps $N$ times. In each iteration, during the first step, set the value $x_k$ as the abscissa of the intersection point from the previous iteration.

Seeking the abscissa of intersection point in the *k-th* iteration is equivalent to calculating the following sequence

$$x_{k+1} = g(x_k) = x_k - p_2(x_k)/p_2'(x_k) = x_k - ((x_k - \bar{x})^2 + s^2)/(2(x_k - \bar{x}))\ (03)$$

For previous purpose, we can write the following algorithm

**Algorithm** (1)

```
1   Set the values of a, b, c (polynomial parameters), N (total number of iteration)
2   Create an array x with size N and set the initial value x[1]
3   if (b² − 4ac < 0) then :
4       x̄ ← -b/(2a)
5       s ← √((4ac − b²)/4a²)
    To evaluate the p₂(x) at x
6       Function Pol(x, x̄, s) do
7           poly ← a((x − x̄)² + s)
8       Return poly
    To evaluate the p₂'(x) at x
9       Function deriPol(x, x̄, s) do
10          polde ← (2a(x − x̄))
11      Return polde

12      epsilon ← 10⁻⁹
13      for k = 1, N − 1 do
14          If x[k] ≠ x̄ then
15              x[k + 1] ← x[k] − Pol(x, x̄, s) /deriPol(x, x̄, s)
16          Else
17              x[k] ← x̄ + epsilon
18              x[k + 1] ← x[k] − Pol(x, x̄, s) /deriPol(x, x̄, s)
19
20          End if
21
22      End for
23  End if
24  Return x
```

Algorithm (1) is just minor modification of Newton's method algorithm. Additionally, when $x_k$ equals 0, we have to use a small number for $x_k$ to avoid division by zero.

We mention here that for the function $g: \mathbb{R}^* \rightarrow \mathbb{R}$, and then the abscissa of the intersection point can be any real number from the set $\mathbb{R}$.

## 3. Relative frequency density ( RFD)

In this section, we present the histograms illustrating the relative frequency density of the algorithm (1) outputs using different polynomials (different values of $s$ and $\bar{x}$ ) and for different values of $N$

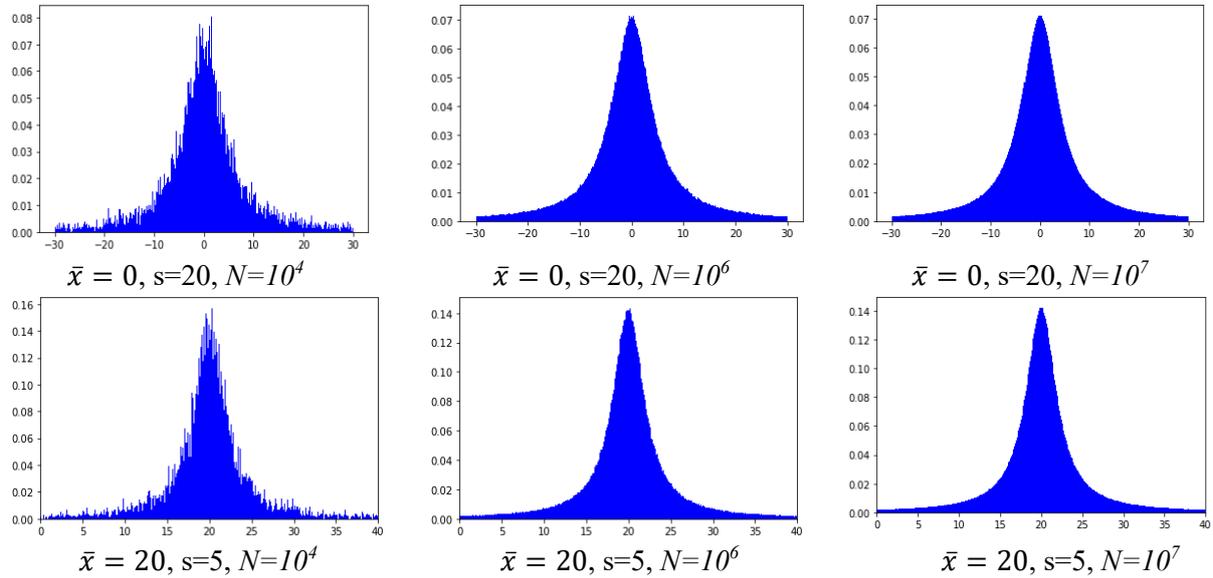

$\bar{x} = 0$, s=20, $N=10^4$  $\bar{x} = 0$, s=20, $N=10^6$  $\bar{x} = 0$, s=20, $N=10^7$

$\bar{x} = 20$, s=5, $N=10^4$  $\bar{x} = 20$, s=5, $N=10^6$  $\bar{x} = 20$, s=5, $N=10^7$

**Figure 2**. histograms of RFDs, the width of bins are fixed at 0.1

As we see in figure 2 histograms are symmetrical and centred around $\bar{x}$ and exhibit heavy tail, another observation is that if the sample size is large, the histogram of relative frequency density converges to a well-defined mathematical function. We can mention here, that $\bar{x}$ is median and it is also the abscissa of turning point of parabola.

In order to understand how we can change the histograms shapes by changing parameters, we use in algorithm (1) different polynomials $p_2(x) = ax^2 + c$ which belongs to a set $F$ and we set $\bar{x} = 0, N=10^7$

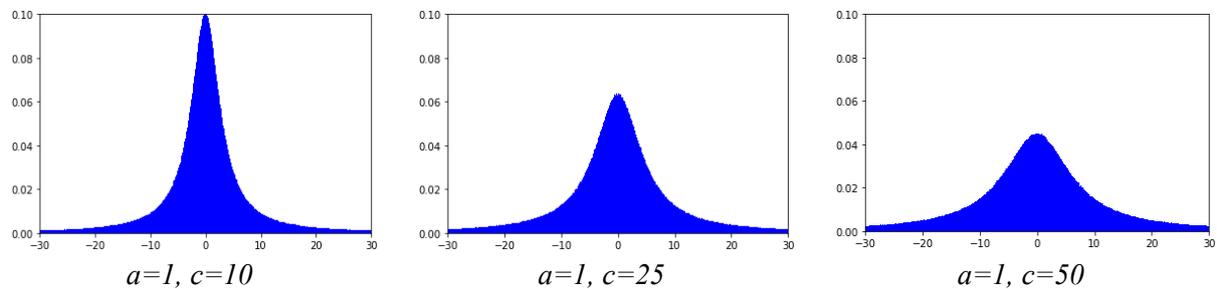

a=1, c=10  a=1, c=25  a=1, c=50

**Figure 3. histograms of RFDs of algorithm (1) outputs for different values of $c$**

If parameter $a$ is fixed and parameter $c$ is varied, as shown in the figure 3, when $c$ is increased, the histogram of relative frequency density becomes more spread out.

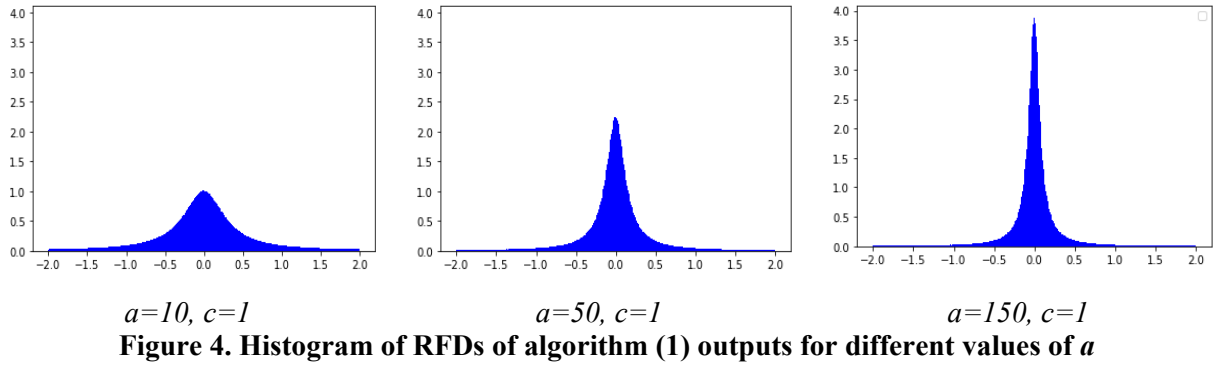

| a=10, c=1 | a=50, c=1 | a=150, c=1 |

**Figure 4. Histogram of RFDs of algorithm (1) outputs for different values of *a***

If parameter *c* is fixed and parameter *a* is varied, we can observe in the Figure 4, the histograms of relative frequency becomes more tightly clustered about the median $\bar{x}$ with increasing the value of *a*. In this particular case, the maximum relative frequency density is function of *a/c*.

## 4. estimated mean and variance

In the following table, we attempt to estimate the mean $M = \sum_{i=1}^{N} x_i/N$ (05) and the variance $var = \frac{1}{(N-1)} \sum_{i=1}^{N}(x_i - M)^2$ (06) of algorithm (1) outputs for parameter $s=1$, $\bar{x} = 0$, $x_1 = 10$

| Sample size (N) | $10^2$ | $10^3$ | $10^4$ | $10^5$ | $10^6$ | $10^7$ |
|---|---|---|---|---|---|---|
| The estimated value of mean (*M*) | 1.7 | 1.84 | -0.12 | -0.97 | -2.72 | -1.28 |
| The estimated value of variance (*var*) | 58.2 | 318 | 2444 | 71340 | 2143714 | 8769028 |

**Table 1.** estimated mean and variance

From previous table, the estimated mean don't tend toward fixed value and the estimated variance values increase as *N* becomes larger. Let's consider the output of algorithm (1) *X* is continuous random variable. Based on preceding information, it is obvious that *X* does not follow the Gaussian distribution. In next section, we construct a solid mathematical proof of the exact formula for the probability density function of *X*.

## 5. Derivation of probability density function and Cumulative distribution function

### 5.1 probability density function *(PDF)*

We introduce the following lemma that can help us to derive *PDF*

**Lemma 1**: $\forall \mu, \lambda \in \mathbb{R}$, If the output of algorithm (1) $x_{k+1} \in [\mu, \lambda]$ then the previous output $x_k$ must be in $[\alpha, \beta]$ or in $[\gamma, \eta]$

Such that $\begin{cases} \alpha = \mu + \sqrt{(\mu - \bar{x})^2 + s^2} \\ \beta = \lambda + \sqrt{(\lambda - \bar{x})^2 + s^2} \end{cases}$ (07) and $\begin{cases} \gamma = \mu - \sqrt{(\mu - \bar{x})^2 + s^2} \\ \eta = \lambda - \sqrt{(\lambda - \bar{x})^2 + s^2} \end{cases}$ (08)

**Proof**

In the domain $x > \bar{x}$, $g(x) = x - \frac{(x-\bar{x})^2 + s}{2(x-\bar{x})} = y$ has inverse function $g^{-1}(y) = \varphi_1(y) = y + \sqrt{(y - \bar{x})^2 + s^2}$ (09)

In the domain $x < \bar{x}$, $g(x)$ has inverse function $g^{-1}(y) = \varphi_2(y) = y - \sqrt{(y - \bar{x})^2 + s^2}$ (10)

The value of previous output $x_{k+1}$ must be equals $x_k^+ > \bar{x}$ or $x_k^- < \bar{x}$ such that

$\begin{cases} x_k^+ = \varphi_1(x_{k+1}) = x_{k+1} + \sqrt{(x_{k+1} - \bar{x})^2 + s^2} \\ x_k^- = \varphi_2(x_{k+1}) = x_{k+1} - \sqrt{(x_{k+1} - \bar{x})^2 + s^2} \end{cases}$ (11)

The function $\varphi_1(y)$ is continuous and differentiable on $\mathbb{R}$ and it is monotonically increasing because $\forall y \in R$, $\varphi_1'(y) > 0$; So, If $x_{k+1} \in [\mu, \lambda]$ then $\varphi_1(x_{k+1}) \in [\varphi_1(\mu), \varphi_1(\lambda)]$ which means $x_k^+ \in [\alpha, \beta]$

Also the function $\varphi_2(y)$ is continuous and differentiable on $\mathbb{R}$ and it is monotonically decreasing because $\forall y \in R$, $\varphi_2'(y) < 0$; So, If $x_{k+1} \in [\mu, \lambda]$ then $\varphi_2(x_{k+1}) \in [\varphi_2(\mu), \varphi_2(\lambda)]$ which means $x_k^- \in [\gamma, \eta]$.

We can understand the previous Lemma from graphical perspective figure 5, if any tangent line of parabola which has a slope $p'(x_k)$ verify the following inequalities $p'(\alpha) < p'(x_k) < p'(\beta)$ or $p'(\gamma) < p'(x_k) < p'(\eta)$, this tangent line must intersect with x-axis in point $(x_{k+1}, 0)$ and $x_{k+1}$ must be in $[\mu, \lambda]$.

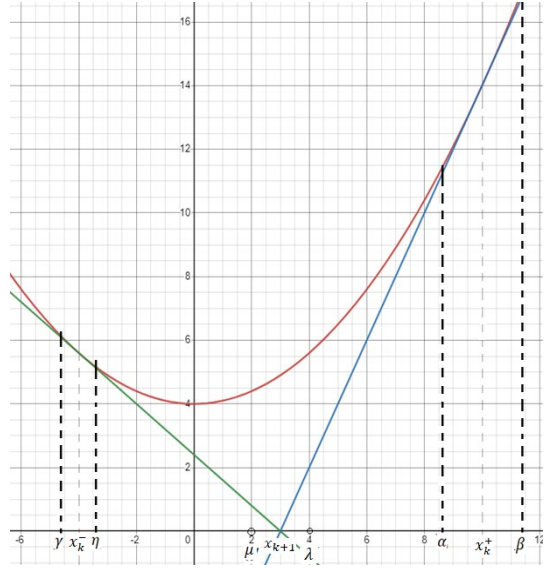

**Figure 5. Understanding the lemma from graphical perspective**

suppose that we have $N$ outputs of Algorithm (1), if we detect $n_1$ outputs belongs to $[\alpha, \beta]$ and if we detect $n_2$ outputs belongs to interval $[\gamma, \eta]$ then according to lemma 1, the number of outputs must be detected in $[\mu, \lambda]$ equal $n_1 + n_2$. In terms of probability, $P(\mu < X < \lambda) = P(\alpha < X < \beta) + P(\gamma < X < \eta)$. The requested formula of probability density function of $X$ must be satisfying the following condition:

$$\int_\mu^\lambda \rho_X(x)\,dx = \int_\alpha^\beta \rho_X(x)\,dx + \int_\gamma^\eta \rho_X(x)\,dx \quad (12)$$

In figure 4, the histograms clearly exhibit heavy tails, and as the value of $x$ increases, the relative frequency density decreases with a polynomial ratio. This behaviour has led to the formulation of a suggested probability density function, which can be represented by the following formula:

$$\rho_X(x) = C/|p_2(x)| = C/|a((x - \bar{x})^2 + s^2)| \quad (13).$$

The absolute value is added because we can use also in process described in section 1 the polynomial $p_2(x) < 0$ see figure1

The *PDF* is positive and continuous on $\mathbb{R}$ and by applying this condition $\int_{-\infty}^{\infty} \rho_X(x)\,dx = 1$, we can find the formula of the normalization constant $C = \frac{1}{2\pi}\sqrt{D}$.

At this point, we have obtained the definitive expression of *PDF* $\rho_X(x) = \dfrac{s/\pi}{((x-\bar{x})^2 + s^2)}$

(14) it is famous in the fields of statistics and probability [04], often referred to as the Cauchy distribution with location parameter $\bar{x} = \dfrac{-b}{2a}$ and scale parameter $s = \sqrt{\dfrac{4ac-b^2}{4a^2}} > 0$.

**5.2 Cumulative distribution function**

The exact formula of *CDF* can be calculated by $F(x) = P(X < x) = \int_{-\infty}^{x} \rho(x) dx$

$$F(x) = \frac{1}{2} + \frac{1}{\pi} arctg\left(\frac{x-\bar{x}}{s}\right) = \frac{1}{2} + \frac{1}{\pi} arctg\left((x + \frac{b}{2a})\sqrt{4a^2/D}\right) \text{ (15)}$$

Having the explicit formula of *CDF*, we can verify the condition (12) is satisfied

The RHS of condition (12) equals $\int_{\mu}^{\lambda} \rho_X(x)\, dx = \dfrac{1}{\pi}\left[arctg\left(\dfrac{\lambda-\bar{x}}{s}\right) - arctg\left(\dfrac{\mu-\bar{x}}{s}\right)\right]$

And LRS of condition (12) equals

$$\int_{\alpha}^{\beta} \rho_X(x)\, dx + \int_{\gamma}^{\eta} \rho_X(x)\, dx = \frac{1}{\pi}\left[arctg\left(\frac{\beta-\bar{x}}{s}\right) - arctg\left(\frac{\alpha-\bar{x}}{s}\right)\right] + \frac{1}{\pi}\left[arctg\left(\frac{\eta-\bar{x}}{s}\right) - arctg\left(\frac{\gamma-\bar{x}}{s}\right)\right] \text{(16)}$$

If we use the following property of the inverse tangent function

$$arctg(h_1) + arctg(h_2) = arctg\left(\frac{h_1 + h_2}{1 - h_1 h_2}\right) + k\pi \;,\; \forall h_1, h_2 \in R \;,\; k \in N \text{ (17)}$$

Then we can prove the following

$$arctg\left(\frac{\lambda-\bar{x}}{s}\right) = arctg\left(\frac{\beta-\bar{x}}{s}\right) + arctg\left(\frac{\eta-\bar{x}}{s}\right) + k\pi \;;\; \left(\frac{\alpha-\bar{x}}{s}\right)\left(\frac{\gamma-\bar{x}}{s}\right) < 0 \text{ then } k = 0 \text{ (18)}$$

$$arctg\left(\frac{\mu-\bar{x}}{s}\right) = arctg\left(\frac{\alpha-\bar{x}}{s}\right) + arctg\left(\frac{\gamma-\bar{x}}{s}\right) + k\pi \;;\; \left(\frac{\beta-\bar{x}}{s}\right)\left(\frac{\eta-\bar{x}}{s}\right) < 0 \text{ then } k = 0 \text{ (19)}$$

In this point, we present some evidence that probability density function of $X$ is Cauchy distribution:

First, the proposed *PDF* (14) is the correct choice because it is continuous, symmetrical, unimodal, and satisfied the condition (12).

Second, Cauchy distribution $C(\bar{x}, s)$ fit very well the histogram of relative frequency density of sample generated by algorithm (1) with parameter $\bar{x}, s$ see the following figure 6

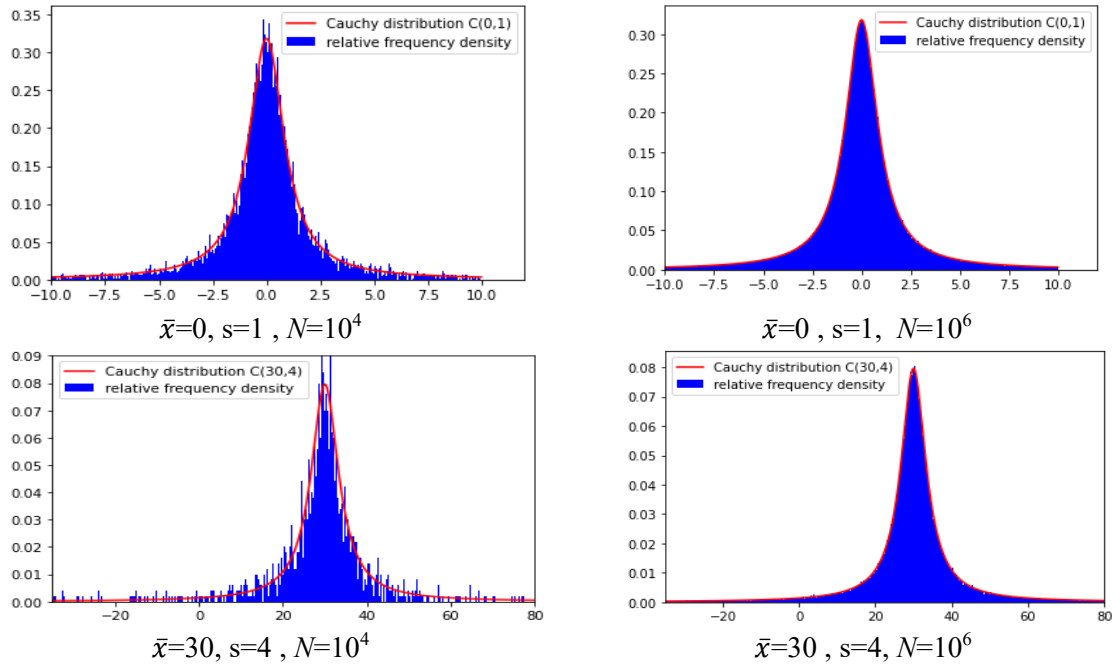

$\bar{x}=0$, s=1, $N=10^4$  $\bar{x}=0$, s=1, $N=10^6$

$\bar{x}=30$, s=4, $N=10^4$  $\bar{x}=30$, s=4, $N=10^6$

**Figure 6. Comparison between *PDF*s and *RFDs***

Third, the estimated mean and variance do not tend to fixed values for large sample sizes as shown in section 4. This indicates that these samples follow a distribution that does not satisfy the law of large numbers, it is legitimate to say that our samples generated by algorithm (01) follow Cauchy distribution because this distribution has no mean or variance and does not satisfy the law of large numbers

Fourth, if *b=0* and $\bar{x} = 0$ then $\max(\rho_X(x)) = \frac{1}{\pi}\sqrt{a/c}$ (20) this information is confirmed by histograms in section 3 see figure 3 and 4.

## 6. Statistical tests

If we use the polynomial $p(x) = x^2 + 1$ in previous process described in section 1, and start with initial value $(x_1 - \bar{x})/s$, then searching for the abscissa of intersection point in the *k-th* iteration is analogous to computing the following sequence $y_{k+1} = y_k - (y_k^2 + 1)/(2y_k) = 0.5(y_k - 1/y_k)$ with $y_1 = (x_1 - \bar{x})/s$. We can also use the following algorithm for the same purpose

**Algorithm (2)**

```
25  Set the value of x̄, s and create an array y with size N
26  Set the initial value y[1] = (x[1] − x̄)/s
27      epsilon ← 10⁻⁹
28      for k = 1, N-1  do
29          If  y[k] ≠ 0  then
30              y[k + 1] ← 0.5(y[k] − 1/y[k])
31          Else
32              y[k] ← epsilon
33              y[k + 1] ← 0.5(y[k] − 1/y[k])
34          End if
35      End for
36  Return x
```

There is a relation between outputs of algorithm (01) and (02) if we use the same values of inputs $\bar{x}, s, x_1$ because we can prove by induction that the sequence (03) can be written as $x_k = sy_k + \bar{x}$ (21)

If we consider the output of algorithm (2) $Y$ is continuous random variable then $X$ is just a linear function of $Y$ such that $X = sY + \bar{x}$

For statistical testing, we only need to test if the sample generated by algorithm (2) follows the standard Cauchy distribution then by using the previous relation (21) we can conclude that $X$ follow a Cauchy distribution with location parameter $\bar{x}$ and scale parameter $s$

Furthermore, we know if $Y$ is continuous random variable which follows standard Cauchy distribution then $U = 0.5 + tan^{-1}(Y)/\pi$ (22) is random variable which follows a standard uniform distribution [05]. In this context, the following algorithm is proposed

**Algorithm (3)**

```
1   Create an array y and u with size N and set the initial value u[1]
2       y[1] ← tan((u[1] − 0.5)π)
3       epsilon ← 10⁻⁹
4       for k = 1, N-1 do
```

|   |   |
|---|---|
| 5 | If $y[k] \neq 0$ then |
| 6 | $y[k+1] \leftarrow 0.5(y[k] - 1/y[k])$ |
| 7 | $u[k+1] \leftarrow 0.5 + \arctan(y[k+1])/\pi$ |
| 8 | Else |
| 9 | $y[k] \leftarrow epsilon$ |
| 10 | $y[k+1] \leftarrow 0.5(y[k] - 1/y[k])$ |
| 11 | $u[k+1] \leftarrow 0.5 + \arctan(y[k+1])/\pi$ |
| 12 | End if |
| 13 | End for |
| 14 | Return $u$ |

In the end, if our sample generated by algorithm (3) follows a standard uniform distribution and is accepted, then we can directly conclude that the sample generated by algorithm (1) or (2) has Cauchy distribution.

### 6.1. Mean and Standard deviation

The histograms of relative frequency density of algorithm (3) outputs are shown in following figure

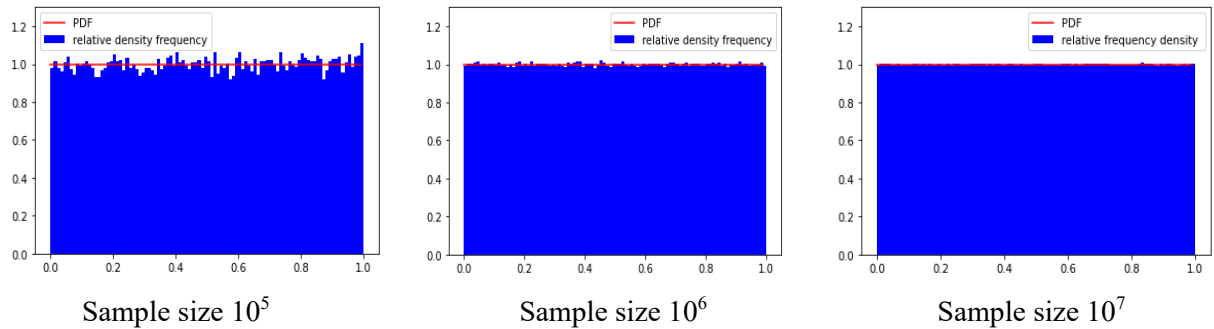

Sample size $10^5$    Sample size $10^6$    Sample size $10^7$

**Figure 7. Histograms of *RFD* of algorithm (3) outputs**

From figure 7, it appears that as the size sample increases, the shape of histograms converge to a plateau. This indicates that all values within the range [0, 1] are equally likely. The values of estimated mean and standard deviation for different sequences generated by algorithm (3) are in good agreement with the theoretical mean and standard deviation of standard uniform distribution.

### 6.2. Goodness-of-fit tests 
are statistical tests used to assess how well an observed sample of data fits a particular theoretical probability distribution. In this paper, we compare the empirical distribution function (*EDF*) derived by random samples generated by Algorithm (3) to the theoretical distribution function (*CDF*) of the standard uniform distribution.

Let's define our null hypothesis ($H_0$) and alternative hypothesis $H_1$

$H_0$: the sequence of numbers generated by algorithm (3) has standard uniform distribution.

*H₁*: the sequence of numbers generated by algorithm (3) does not have the standard uniform distribution.

In this context, to test this hypothesis, various statistical tests are performed, including the Cramér-Von Mises, Kolmogorov-Smirnov, Anderson-Darling, and Watson statistics. The theoretical basis of these tests is discussed in books [06].

The level of significant $\alpha$ is fixed at 0.01 and we use a large sample with a size of $10^6$, if the specific test returns *p*-value $\geq \alpha$ then the null hypothesis is accepted.

| test | Anderson-Darling | Kolmogorov-Smirnov | Watson statistics | Cramér-Von Mises |
|---|---|---|---|---|
| ***p-value*** | 0.8 | 0.98 | 0.99 | 0.93 |

**Table 2.** Values of *p_value* for different goodness-fit-tests

From the values in table 2, Algorithm (3) has successfully passed the main tests and the null hypothesis is accepted. We confirm once more that Algorithm (02) generates samples following a standard Cauchy distribution.

Finally, with evidence presented in previous sections, we are able to answer the following question: in the iterative process described in section 1, what is the probability of the intersection event happening in interval [*x1, x2*]?

The answer is $p(x_1 < X < x_2) = \frac{1}{\pi}\left[arctg\left(\left(x_2 + \frac{b}{2a}\right)\sqrt{\frac{4a^2}{4ac-b^2}}\right) - arctg\left(\left(x_1 + \frac{b}{2a}\right)\sqrt{\frac{4a^2}{4ac-b^2}}\right)\right]$ (23).

## 7. Distribution of distances between two successive intersections

To create another new process, we adopt all the steps described in section 1 and additionally include a final step involving the calculation of the distance between the previous intersection and the current intersection $r_k = |x_{k+1} - x_k|$ (24)

For the previous purpose, we rewrite a new algorithm (4) by modifying Algorithm (1) in a loop as follows:

**In loop**

| | |
|---|---|
| 1 | for $k = 1, N-1$ do |
| 2 | If $x[k] \neq \bar{x}$ then |
| 3 | $x[k+1] \leftarrow x[k] - \text{Pol}(x, \bar{x}, s)/\text{deriPol}(x, \bar{x}, s)$ |
| 4 | Else |
| 5 | $x[k] \leftarrow \bar{x} + epsilon$ |
| 6 | $x[k+1] \leftarrow x[k] - \text{Pol}(x, \bar{x}, s)/\text{deriPol}(x, \bar{x}, s)$ |
| 7 | |
| 8 | End if |
| 9 | $r[k+1] \leftarrow abs(x[k+1] - x[k])$ |
| 10 | End for |

The histogram of relative frequency density of distances generated by algorithm (4) for different input values of $\bar{x}$ and $s$ is shown in the following figure

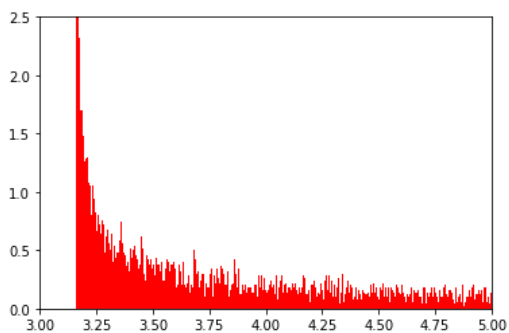 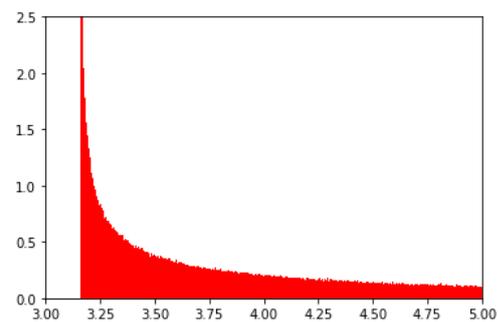

$N=10^4, \bar{x} = 0, s = 3.16$  $\qquad\qquad$  $N=10^6, \bar{x} = 0, s = 3.16$

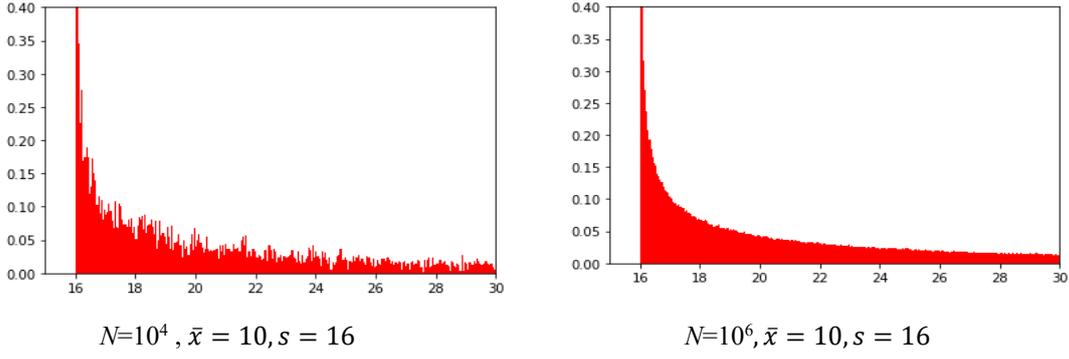

$N=10^4$, $\bar{x}=10, s=16$          $N=10^6, \bar{x}=10, s=16$

**Figure 8. Histograms of relative frequency density of distances**

As we observe, the histograms exhibit decay as the distance value increases. Moreover, as the sample size increases, we notice that the histograms of relative frequency density converge to well-defined mathematical functions. Let's consider distance calculated by algorithm (4) as a continuous random variable denoted $R$ and seek its probability density function.

For the first iteration, the distance $r_1 = |x_2 - x_1| = \left| \dfrac{-((x_1 - \bar{x})^2 + s^2)}{2(x_1 - \bar{x})} \right|$ (25),

for $k$-th iteration the distance $r_k = |x_{k+1} - x_k| = \left| \dfrac{-((x_k - \bar{x})^2 + s^2)}{2(x_k - \bar{x})} \right|$ (26)

This leads us to conclude that the random variable $R$ is simply a nonlinear change of Cauchy random variable $X$ such that $R = h(X) = \left| \dfrac{-((X - \bar{x})^2 + s^2)}{2(X - \bar{x})} \right|$ (27)

The function $h(X)$ is differentiable and strictly monotone on the following intervals $I_1 = ]-\infty, \bar{x} - s]$; $I_2 = [\bar{x} - s, 0[$; $I_3 = ]0, \bar{x} + s]$; $I_4 = [\bar{x} + s, \infty[$

At this stage, we can apply the method of transformations to derive a probability density function for random variable $R$

$\rho_R(r)dr = \sum_{i=1}^{4} \rho_X(x_i)/|h'(x_i)|$ (28), where $x_i$ are real solutions of $h(x) = r$ in each interval $I_i$

If $x < \bar{x}$ then $r = h(x) = \dfrac{-((x - \bar{x})^2 + s^2)}{2(x - \bar{x})}$ (29) and Solutions in this case are

$\begin{cases} x_1 = \bar{x} - r - \sqrt{r^2 - s^2} \\ x_2 = \bar{x} - r + \sqrt{r^2 - s^2} \end{cases}$ (30)

In this case $h'(x) = \dfrac{-(x - \bar{x})^2 + s^2}{2(x - \bar{x})}$ (31) and the local minimum is $h(\bar{x} - s) = r_{min} = s$ because $h'(\bar{x} - s) = 0$ and $h''(\bar{x} - s) > 0$

If $x > \bar{x}$ then $r = h(x) = ((x - \bar{x})^2 + s^2)/2(x - \bar{x})$ (32) and Solutions in this case are

$$\begin{cases} x_3 = \bar{x} + r - \sqrt{r^2 - s^2} \\ x_4 = \bar{x} + r + \sqrt{r^2 - s^2} \end{cases} (33)$$

In this case $h'(x) = (x - \bar{x})^2 - s^2 / 2(x - \bar{x})$ (34) and the local minimum is $h(\bar{x} + s) = r_{min} = s$ because $h'(\bar{x} + s) = 0$ and $h''(\bar{x} + s) > 0$.

Finally, the formula for the probability density function of $R$ is deduced and given by

$$\rho_R(r) = \frac{4s}{\pi} \left( \frac{1}{(L_1^2 + s^2)} \left| \frac{(L_1)^2}{(L_1)^2 - s^2} \right| + \frac{1}{(L_2^2 + s^2)} \left| \frac{(L_2)^2}{(L_2)^2 - s^2} \right| \right) (35), \text{ for } r \in ]s, +\infty[$$

Such that $L_1 = r + \sqrt{r^2 - s^2}$ and $L_2 = r - \sqrt{r^2 - s^2}$

In the following figure, we show a comparison between the previous probability density function and the relative frequency density of distances.

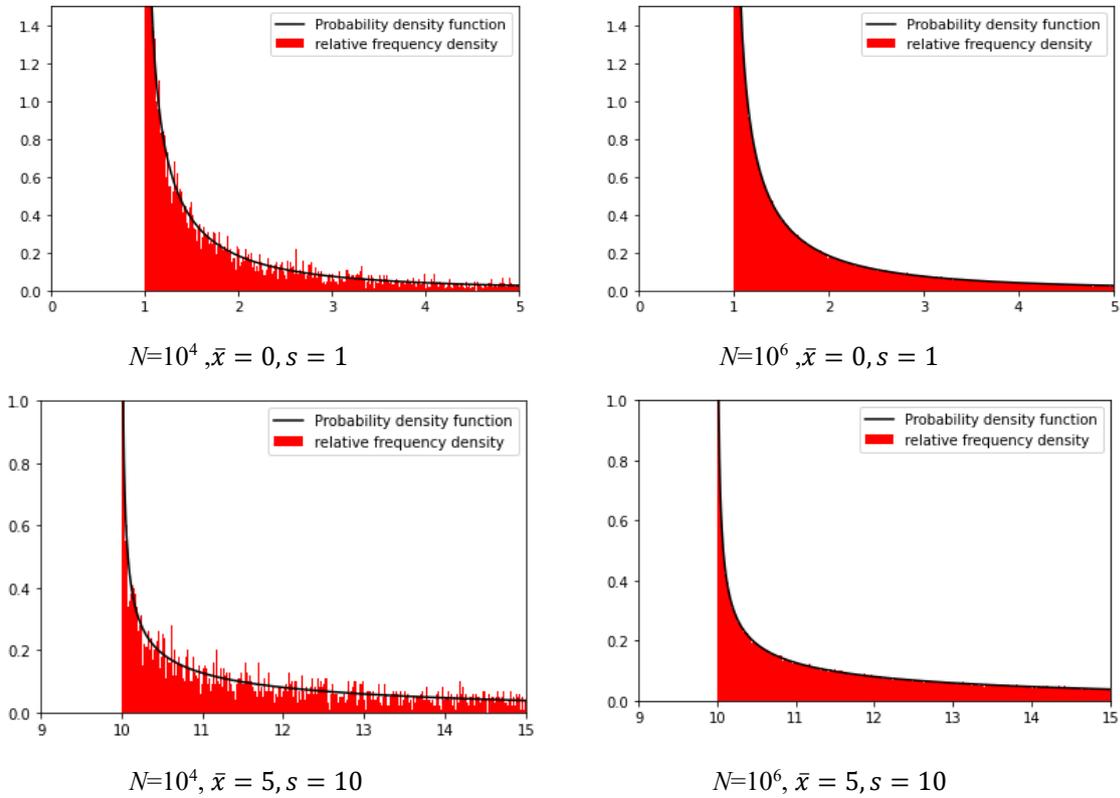

$N=10^4, \bar{x} = 0, s = 1$          $N=10^6, \bar{x} = 0, s = 1$

$N=10^4, \bar{x} = 5, s = 10$          $N=10^6, \bar{x} = 5, s = 10$

**Figure 9. comparison between probability density function $\rho_R(r)$ and relative frequency density**

The probability density function $\rho_R(r)$ fits very well with histogram of relative frequency density of distances generated by algorithm (4) with parameters $\bar{x}, s$. This provides further evidence that $X$ is just Cauchy random variable.

**Conclusion and Perspectives**

In this paper, the Newton's method is demonstrated in a specific case where it can generate a sample following the Cauchy distribution. The theoretical proof mentioned in section 5 and goodness-of-fit tests confirm this result.

In the field of numerical analysis, our primary focus often revolves around understanding the convergence behaviour of various types of sequences. We highly recommend delving into the statistical properties of sequences like (03) in case when we use polynomials of degrees higher then two that do not have roots and we also encourage deriving the probability density function for their multimodal distribution.